\documentclass[twocolumn,preprintnumbers,amsmath,amssymb]{revtex4}

\usepackage{bm}%
\usepackage{graphicx, color}
\usepackage{comment}
\expandafter\ifx\csname package@font\endcsname\relax\else
 \expandafter\expandafter
 \expandafter\usepackage
 \expandafter\expandafter
 \expandafter{\csname package@font\endcsname}%
\fi

\begin{document}

\title{Electronic and magnetic properties of off-stoichiometric Co$_\mathrm{2}$Mn$_{\beta}$Si/MgO interfaces studied by x-ray magnetic circular dichroism}%

\author{V. R. Singh}
\thanks{\it Presently: Department of Physics, Boston University, 590 Commonwealth Avenue, Boston, Massachusetts 02215, United States}

\author{V. K. Verma}
\thanks{\it Presently: Department of Physics, Madanapalle Institute of Technology and Science,
Angallu, Madanapalle 517325, AP, India}

\author{K. Ishigami}
\author{G. Shibata}
\author{A. Fujimori}
\email{fujimori@wyvern.phys.s.u-tokyo.ac.jp}
\affiliation{Department of Physics, University of Tokyo, Bunkyo-ku, Tokyo 113-0033, Japan}

\author{T. Koide}
\affiliation{Photon Factory, IMSS, High Energy Accelerator Research Organization, Tsukuba, Ibaraki 305-0801, Japan}

\author{Y. Miura}
\thanks{\it Presently: Mathematical and Physical Science, Kyoto Institute of Technology, Matsugasaki, Kyoto 606-8585, Japan}

\author{M. Shirai}
\affiliation{Research Institute of Electrical Communication, Tohoku University, Katahira 2-1-1, Aoba-ku, Sendai 980-8577, Japan}

\author{T. Ishikawa}
\thanks{\it Corporate R \& D Center, Toshiba Corp., Kawasaki 212-8582, Japan}

\author{G. f. Li}
\thanks{\it School of Electronics and Information, Northwestern Polytechnical University, Xi'an 710072, China}

\author{M. Yamamoto}
\affiliation{Division of Electronics for Informatics, Hokkaido University, Sapporo 060-0814, Japan}

\date{\today}

\begin{abstract}

We have studied the electronic and  magnetic  states of Co and Mn atoms at the interface of the  Co$_\mathrm{2}$Mn$_{\beta}$Si (CMS)/MgO ($\beta$=0.69, 0.99, 1.15 and 1.29) magnetic tunnel junction (MTJ) by means of x-ray magnetic circular dichroism. In particular, the Mn composition ($\beta$) dependences of the Mn and Co magnetic moments were investigated. The experimental spin magnetic moments of Mn, $m_\mathrm{spin}$(Mn), derived from XMCD weakly decreased with increasing Mn composition $\beta$  in going from Mn-deficient to Mn-rich  CMS films. This behavior was explained by first-principles calculations based on the antisite-based site-specific formula unit (SSFU) composition model, which assumes the formation of only antisite defect, not vacancies, to accommodate off-stoichiometry. Furthermore, the experimental spin magnetic moments of Co, $m_\mathrm{spin}$(Co), also weakly decreased with increasing Mn composition. This behavior was consistently explained by the antisite-based SSFU model, in particular, by the decrease in the concentration of Co$_\mathrm{Mn}$ antisites detrimental to the half-metallicity of CMS with increasing $\beta$. This finding is consistent with the higher TMR ratios which have been observed for CMS/MgO/CMS MTJs with Mn-rich CMS electrodes.  

\end{abstract}


\maketitle

\section{Introduction}

\ \ \ A highly efficient spin-polarized electron source is a key element for spintronic devices including tunnel magnetoresistance (TMR) devices \cite{1,2,3,4,5,6}. Half-metallic ferromagnets (HMFs) are one of the most suitable materials for spin-polarized electron sources because they provide 100\% spin polarization at the Fermi level ($E_\mathrm{F}$) \cite{1,2,3,4,5,6}. Among the HMFs, Co-based Heusler alloys are especially promising due to their high Curie temperatures, which are well above room temperature  (RT) \cite{6b}. Recently, extensive studies have been conducted to apply Co-based Heusler alloy (Co$_\mathrm{2}YZ$, where Y is usually a transition metal and Z is a main group element) thin films sandwiching a MgO tunnel barrier to spintronic devices and the interfaces between the Huesler alloys and MgO have become an important subject  because the TMR ratio is sensitive to the interfaces \cite{1,2,3,4,5,6}.

\ \ \ Off-stoichiometry and associated structural defects are expected, to various degrees, in Co$_\mathrm{2}YZ$ thin films prepared by magnetron sputtering or molecular beam epitaxy methods \cite {3}. The defects are expected to profoundly influence the half-metallic behavior and hence the TMR characteristics \cite{3, 7,8,9,10}. The effect of defects in Co$_\mathrm{2}YZ$ introduced by off-stoichiometry on the spin-dependent electronic structure has been investigated theoretically \cite{7,8,9,10} as well as experimentally \cite{2,3,4,5,6,11,12}. Picozzi {\it et al.}, theoretically predicted that Co$_\mathrm{Mn}$ antisites in Co$_\mathrm{2}$MnSi and Co$_\mathrm{2}$MnGe, where a Mn site is replaced by a Co atom, cause minority-spin in-gap states near $E_\mathrm{F}$  and thus are detrimental to the half-metallicity of Co$_\mathrm{2}$MnSi and Co$_\mathrm{2}$MnGe \cite{7}. Miura {\it et al.} \cite{8} theoretically predicted that disorder between Co sites and Cr sites in Co$_\mathrm{2}$CrAl leads to a significant reduction in the spin polarization at $E_\mathrm{F}$, while the disorder between Cr sites and Al sites does not significantly decrease the spin polarization at $E_\mathrm{F}$. Yamamoto and co-workers \cite{3,4,5,12} have made systematic studies of the effect of non-stoichiometry in the Heusler alloy Co$_\mathrm{2}$Mn$_{\beta}$Si (CMS) on the spin-dependent tunneling characteristics of CMS/MgO/CMS magnetic tunnel junctions (CMS MTJs) with  Co$_\mathrm{2}$Mn$_{\beta}$Si$_{\gamma}$ electrodes having various Mn compositions $\beta$  and a fixed Si composition $\gamma$ and demonstrated high TMR ratios of 1995\% at 4.2 K and 354\% at RT for CMS MTJs with Mn-rich CMS electrodes \cite{5}.  On the other hand, identically fabricated Ge-deficient Co$_\mathrm{2}$Mn$_{\beta}$Ge (CMG)-based CMG/MgO/CMG MTJs (CMG MTJs) with Co$_\mathrm{2}$Mn$_{\beta}$Ge$_\mathrm{0.38}$  electrodes having various $\beta$ ranging from 0.67 to 1.40 showed lower TMR ratios of 650\% at 4.2 K and 220\% at RT with Mn-rich CMG electrodes \cite{4}.  In this context, further understanding of the effect of structural defects associated with non-stoichiometry on the electronic and magnetic properties in Co-based Heusler-alloy electrodes, particularly in the interfacial region between the Co$_\mathrm{2}$Mn$_{\beta}$Si electrode and the MgO tunnel barrier are important for extending the application of Heusler alloy films to spintronic devices.

\ \ \ Recently, x-ray absorption spectroscopy (XAS) and x-ray magnetic circular dichroism (XMCD) have been proved to be effective techniques for obtaining microscopic information about the element-specific electronic and magnetic states in the interfacial region of MTJs \cite{1,2, 6,13,14,15,16,17,18}.  In a more recent work \cite{19}, we performed XAS and XMCD studies on Co$_\mathrm{2}$Mn$_{\beta}$Ge$_\mathrm{0.38}$ (CMG) thin films facing an MgO barrier using the surface/interface-sensitive total electron yield (TEY) mode and found that charges and spin moment on Mn atoms are reduced because of the delocalization of Mn 3$d$ electrons with increasing $\beta$ and that the Co spin magnetic moments for all the samples are increased with decreasing $\beta$ because Co$_\mathrm{Mn}$  antisite has a larger magnetic moment than that of Co atoms at the regular sites but reduces the spin polarization at the Fermi level.

\ \ \ In this work, we have extended the approach applied to CMG to Co$_\mathrm{2}$Mn$_{\beta}$Si (CMS) facing an MgO barrier, which exhibits higher TMR ratios than CMG/MgO MTJs. Density functional calculations were also made for a wide range of nonstoichiometry corresponding to the real material CMS with 0.69 $< \beta  <$ 1.29 which exceeds the range studied by Piccozi {\it et al.}\cite{7} where the properties of a single antisite defect are dominant.

 \section{Methods}

 \ \ \ The samples which we studied had the following layer structures: (from the substrate side) MgO buffer layer (10 nm)/CMS (30nm)/MgO barrier (2 nm) /AlO$_x$  (1 nm) cap, grown on an MgO(001) single-crystal substrate. The preparation of the samples is described in detail elsewhere \cite{3,4}. Each sample layer was successively deposited in an ultrahigh vacuum chamber with a base pressure $\sim$ 6 $\times$ 10$^{-8}$ Pa through magnetron sputtering for CMS and electron beam evaporation for MgO.  CMS thin films having film compositions of Co$_\mathrm{2}$Mn$_{\beta}$Si$_\mathrm{1}$ with various Mn compositions $\beta$ ranging from 0.69 to 1.29 were prepared by co-sputtering from a nearly stoichiometric CMS target and a Mn target and
 the films deposited on the MgO buffer were subsequently annealed in situ at 600$^{\circ}$C for 15 min.  The transmission electron microscopy (TEM) images show very smooth and abrupt interfaces and all layers were grown epitaxially \cite{3,4}. It is also clear that they have $L$2$_\mathrm{1}$ structure \cite{3,4}. The XAS and XMCD measurements were performed at BL-16A of Photon Factory (KEK), Japan. The monochromator resolution was $E$/$\Delta E >$  10000, the degree of circular polarization of x-rays was 87\%$\pm$ 4\%, the base pressure of the chamber was about 10$^{-9}$ Torr and the sample temperature was maintained at 300 K. XAS and XMCD spectra were obtained in the total electron yield (TEY) mode without surface preparation. The probing depth of the TEY mode was  $\sim$ 5 nm. XMCD was measured in a magnetic field ($\pm$ 3 T) applied perpendicular to the films at 300 K and 20 K. 
 
 \ \ \ Since Picozzi {\it et al.}'s \cite{7} calculation was for stoichiometries only down to Mn$_\mathrm{0.9}$ and Ge$_\mathrm{0.9}$, we performed the density functional calculations on the basis of the Korringa-Kohn-Rostoker (KKR) method \cite{3,20,21} with the coherent potential approximation (CPA) for a wider range of nonstoichiometries in CMS. We used the generalized gradient approximation \cite{22} for the exchange and correlation term. We adopted the experimental lattice constants obtained for the cosputtered cubic CMS thin films depending on the Mn composition $\beta$ in CMS \cite{3}. The KKR-CPA calculations were based on the site-specific formula unit composition model assuming antisite formation rather than vacancy formation to accommodate nonstoichiometry \cite{3}.

\section{Results and discussion}

\begin{figure}[htbp]
\begin{center}
\includegraphics[width=9cm]{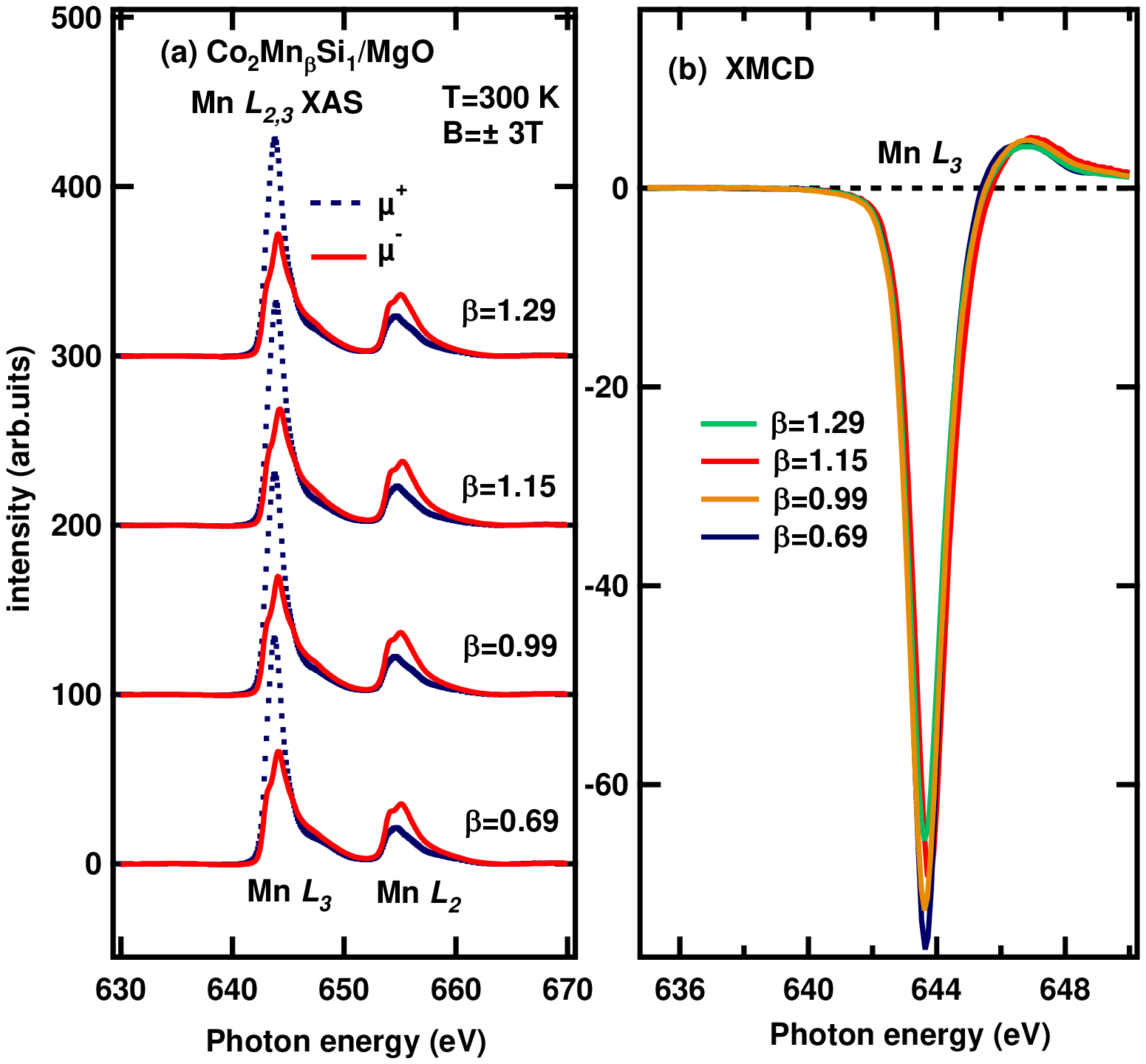}
\caption{(Color online) Mn $L_\mathrm{3,2}$-edge XAS and XMCD of Co$_\mathrm{2}$Mn$_{\beta}$Si samples with various Mn compositions $\beta$ (a) XAS taken at 300 K and B = $\pm$3 T. $\mu^+$ and $\mu^-$ are the absorption coefficients for photon helicity parallel and antiparallel to the Mn 3$d$ majority spin, respectively. (b) Corresponding XMCD spectra.}
\end{center}
\end{figure}

\ \ \ Figure 1(a) shows the photon flux-normalized polarization dependent XAS spectra ($\mu^+$ and $\mu^-$) at the Mn $L_\mathrm{2,3}$- edges (2$p_{3/2,1/2}$ $\rightarrow$  3$d$ absorption). Here, $\mu^+$ and $\mu^-$ stand for the absorption coefficient for the photon helicity parallel and antiparallel to the Mn 3$d$ majority spin. Figure 1(b) displays the Mn $L_\mathrm{2,3}$-edge XMCD ($\Delta\mu$ = $\mu^+$ - $\mu^-$) spectra. In the XAS spectra for Mn compositions $\beta$ from 0.69 to 1.29, a shoulder-like structure was observed on the higher energy side of the Mn $L_\mathrm{3}$ peak, and the Mn $L_\mathrm{2}$ peak was split into a doublet. These features are characteristic of  Co$_\mathrm{2}$MnSi and Co$_\mathrm{2}$MnGe \cite{23}. These features are due to interplay of two effects, namely, the exchange and Coulomb interactions between the core holes and unpaired electrons in the valence band, and strong hybridization between the 3$d$ and surrounding electronic states \cite{23}. Here, Mn atoms were not oxidized in the interfacial region as is clear from the XAS spectrum. The XAS and XMCD intensities at both 20 K and 300 K showed similar $\beta$ dependences, that is, the XMCD intensities decreased with $\beta$ at 20K and 300K.

\ \ \ By applying the sum rules \cite{1,2,19,24,25,26,26A,26B,26C,26D}, we obtained the spin and orbital magnetic moments of the Mn atom from the XAS and XMCD spectra. We assumed the contribution of the dipole operator term $T_z$ in the spin sum rule to be negligibly small because the Mn atom in CMS is located at the high-symmetry $T_d$ site \cite {2}. Since there is an overlapping region between the $L_\mathrm{3}$ and $L_\mathrm{2}$  features due to strong exchange interaction between the Mn 2$p$ core hole and the Mn 3$d$ electrons in Mn  $L_\mathrm{2,3}$-edge XAS and XMCD, we divided the obtained spin-magnetic moment by a correction factor 0.68 given by Teramura {\it et al.}\cite{27}. We assumed, on the basis of band-structure calculation, the 3$d$ hole number ($n_h$) of 4.5 for Mn \cite{17, 28}. As shown in Fig. 2(a), the deduced spin magnetic moment $m_\mathrm{spin}$(Mn) was in the range of 3.02-3.45$\mu_\mathrm{B}$ which decreases slightly as the temperature increases from 20 to 300 K and decreases weakly from $\beta$= 0.69 to 1.0 and then decreases rapidly with increasing $\beta$.  These results are reasonable agreement with the  present KKR-CPA calculation. For $\beta$ =1.29, $m_\mathrm{spin}$(Mn) was nearly the same as the theoretical value of 3.04 $\mu_\mathrm{B}$ \cite{7}. Also the trend of $m_\mathrm{spin}$ (Mn) with $\beta$ was consistent with  present calculation as well as with the calculation by Picozzi {\it et al.} \cite{7}. The Mn orbital magnetic moment $m_\mathrm{orbital}$ (Mn) (not shown) was found to be in the range from $\sim$ 0.30 $\pm$ .004 $\mu_\mathrm{B}$  to $\sim$0.40 $\pm$ .005 $\mu_\mathrm{B}$ for all the samples.  Identically fabricated CMG samples with the film compositions of Ge-deficient Co$_\mathrm{2}$Mn$_{\beta}$Ge$_\mathrm{0.38}$  showed a similar dependence on $\beta$ \cite{19} to the CMS samples with the film compositions of Co$_\mathrm{2}$Mn$_{\beta}$Si$_\mathrm{1}$ as shown in Fig. 2(a) but $m_\mathrm{spin}$(Mn) in the CMG samples is smaller than the CMS \cite{19}. This is due to the difference of the concentration of antisite Mn$_\mathrm{Co}$  in the site-specific formula unit (SSFU) composition model which shows the negative spin magnetic moment arising from its antiferromagnetic coupling with the nearest neighbor Mn atoms at the regular Mn sites \cite{7,19}. Since the CMG samples are strongly Ge-deficient, Mn$_\mathrm{Co}$ antisites exist for a wide range of $\beta$ from 0.67 to 1.60 in the CMG according to the antisite-based SSFU composition model \cite{3,4,19}. This is contrasted with that Mn$_\mathrm{Co}$  antisites exist only for $\beta$ = 1.15 and 1.29 for Co$_\mathrm{2}$Mn$_{\beta}$Si$_\mathrm{1}$. This led to a reduction of the $m_\mathrm{spin}$(Mn) in the CMG samples. Figure 2(a) also shows the dependence of the theoretical $m_\mathrm{spin}$ value for Mn, $m_\mathrm{spin}$(Mn), as a function of $\beta$ in Co$_\mathrm{2}$Mn$_{\beta}$Si$_\mathrm{1}$ obtained by the present KKR-CPA calculations, where the theoretical $m_\mathrm{spin}$(Mn) is an averaged value for those at the different sites (nominal Mn, Si and Co sites) taking into account the concentration of the Mn atoms at the nominal Mn, Si and Co sites.

\begin{figure}[htbp]
\begin{center}
\includegraphics[width=09cm]{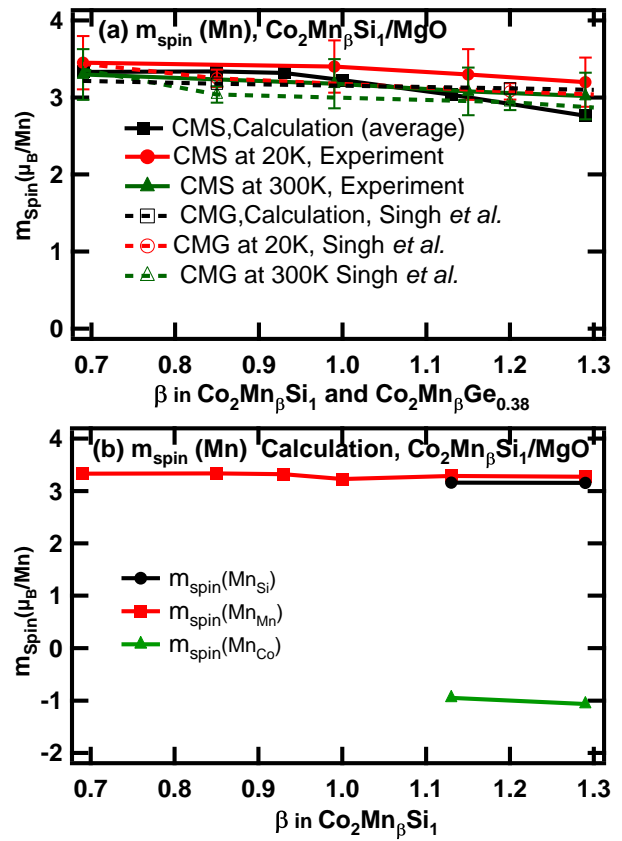}
\caption{(Color online) Mn composition ($\beta$) dependence of the Mn spin magnetic moment deduced from the XMCD data using the sum rules \cite{1,2,19,24,25,26,26A,26B,26C,26D} compared with that given by the present KKR-CPA calculation. The magnetic moments of Mn in CMG by Singh  {\it et al.}\cite{19} are shown by dotted lines. (d) $\beta$ dependence of the Mn spin magnetic moments at different sites deduced from present KKR-CPA calculation.}
\end{center}
\end{figure}

  \ \ \  Figure 2(b) shows the $\beta$ dependence of the Mn spin magnetic moment at different sites deduced from the present KKR-CPA calculation. Here, the magnetic moment of Mn at the regular site,$m_\mathrm{spin}$(Mn$_\mathrm{Mn}$), and that at the Si site, $m_\mathrm{spin}$(Mn$_\mathrm{Si}$), shows a weak decrease with $\beta$, because the Mn atoms at both the regular-site and the Si antisite lose charges in the majority-spin state. Li {\it et al.}\cite{3} showed the local density of states (LDOS) of Mn  in CMS at both the regular-site and the Si antisite and it was found that the LDOSs of Mn  is shifted toward the higher energy side with increasing $\beta$, indicating the reduction of charges of Mn. This can be attributed to the reduction of charge for Mn in terms of the total valence electron number of $Z_t$ of Co$_\mathrm{2}$Mn$_{\beta}$Si$_\mathrm{1}$  provided by the antisite-based SSFU composition model.  According to this model,  $Z_t$-24 for Co$_\mathrm{2}$Mn$_{\beta}$Si$_\mathrm{1}$ decreased with increasing $\beta$ from 5.08 for$\beta$ = 0.69 to 4.93 for $\beta$ = 1.29 [Fig. 6(d) of Ref. \cite{4}]. Similarly,   $Z_t$-24 for Co$_\mathrm{2}$Mn$_{\beta}$Ge$_\mathrm{0.38}$ decreased with increasing $\beta$ from 7.75 for $\beta$ = 0.67 to 6.84 for $\beta$ = 1.60 [Fig. 5(a) of Ref.\cite{4}]. This led to the shift of $E_\mathrm{F}$ toward the lower energy side in the total DOS. Indeed, the $E_\mathrm{F}$ in the theoretically calculated LDOS for Mn at the regular site and the nominal Ge site are  shifted toward the lower energy side with increasing $\beta$  \cite{19}, which is originated from the reduction of $Z_t$. Since the local density of states for Mn in  Co$_\mathrm{2}$Mn$_{\beta}$Ge$_\mathrm{0.38}$  show nearly-half-metallic electronic structures, the reduction of charge is significant in the majority-spin states. This leads to the decrease in the Mn spin moment at both the regular site and the Si site. The experimentally observed rapid decrease of $m_\mathrm{spin}$(Mn) for $\beta$= 1.15 and 1.29 is not due to the $m_\mathrm{spin}$(Mn$_\mathrm{Mn}$) or $m_\mathrm{spin}$(Mn$_\mathrm{Si}$), but to the negative value of $m_\mathrm{spin}$(Mn$_\mathrm{Co}$). According to the site occupation model employed in Table I of Li {\it et al.}\cite{3}, only the $\beta$ = 1.15 and 1.29 samples have a finite number of Mn$_\mathrm{Co}$ antisites and consistently shows a reduced average $m_\mathrm{spin}$(Mn).

\begin{figure}[htbp]
\begin{center}
\includegraphics[width=9cm]{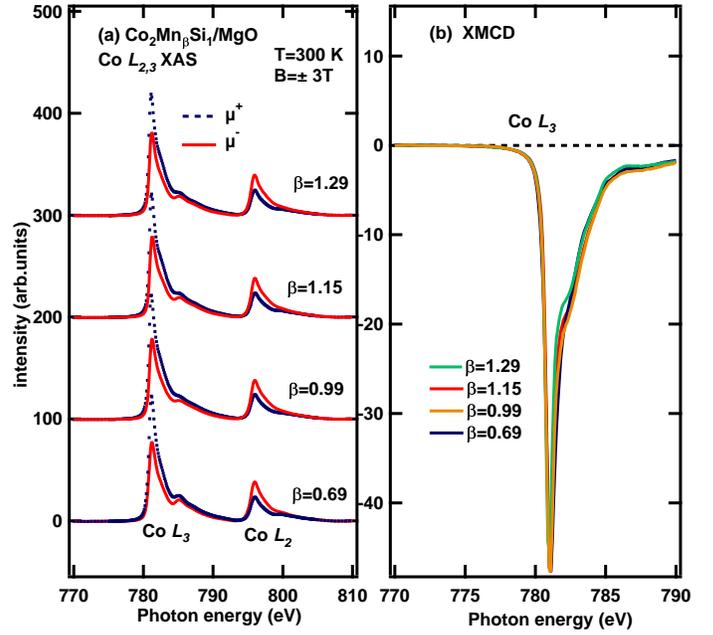}
\caption{(Color online) Co $L_\mathrm{3,2}$-edge XAS and XMCD of Co$_\mathrm{2}$Mn$_{\beta}$Si$_\mathrm{1}$ samples with various Mn composition $\beta$ (a) XAS taken 300 K and B = $\pm$3 T. $\mu^+$ and $\mu^-$ are the absorption coefficients for photon helicity parallel and antiparallel to the Mn 3$d$ majority spin, respectively. (b) Corresponding XMCD spectra.}
\end{center}
\end{figure}

\ \ \ Figure 3(a) shows photon flux-normalized, polarization dependent XAS spectra at the Co $L_\mathrm{3,2}$-edges of CMS. Fig. 3(b) displays the Co $L_\mathrm{3,2}$ XMCD spectra. All the samples showed a shoulder like structure observed in the higher energy region of the Co $L_\mathrm{3}$-edge XAS and XMCD. This feature is common to bulk samples \cite{23}. Although the shoulder at the Co $L_\mathrm{3}$  peak (which is characteristic of Heusler alloys with the $L$2$_\mathrm{1}$  structure) is also observed in the XAS and CMCD spectrum for CMG \cite{19}, though it is not as clear as the shoulder for CMS film. As shown in Fig. 3(b), the XMCD signals are slightly decreasing with $\beta$ values. Co-oxide-like multiplet structures were not observed for any samples, meaning that Co atoms were not oxidized even in the interfacial region, consistent with the literature \cite{1}.

\ \ \ To determine the Co magnetic moment, we applied the sum rules \cite{1,2,19,24,25,26,26A,26B,26C,26D} as in the case of the Mn $L_\mathrm{3,2}$-edges. Since the Co atoms are also in the highly-symmetry $T_d$ crystal field, we again neglect the $T_z$ term in the spin sum rule \cite{2}. Here we assumed, on the basis of a band-structure calculation, a 3$d$ hole number ($n_h$) of 2.2 for Co \cite{17, 28} in the sum-rule analysis.

\ \ \   As shown in Fig. 4(a), we determined the Co spin magnetic moment  $m_\mathrm{spin}$(Co) to be in the range from 1.08  $\mu_\mathrm{B}$ to 1.16 $\mu_\mathrm{B}$ which decreased slightly with increasing temperature (20 to 300 K) and Mn composition ($\beta$). This result is consistent with Li {\it et al.}’s calculation \cite{3}. The decrease of $m_\mathrm{spin}$(Co) with increasing $\beta$ in the experimental result can be understood as being due to the decreasing concentration of Co$_\mathrm{Mn}$ antisites. The orbital magnetic moment $m_\mathrm{orbital}$ (Co) was in the range of 0.3 $\pm$ .001 $\mu_\mathrm{B}$ to 0.1 $\pm$ .003 $\mu_\mathrm{B}$ for all the samples. Identically fabricated CMG samples also showed a similar dependence on $\beta$ to CMS but $m_\mathrm{spin}$(Co) in CMS is smaller than CMG  \cite{19}. This is due to the reduced concentration of Co$_\mathrm{Mn}$ antisites in CMS compared to Ge-deficient CMG. The suppression of Co$_\mathrm{Mn}$ antisites in Mn-rich CMS electrodes will promote towards higher TMR than Ge-deficient CMG \cite{4}.  Thus, our XMCD study suggests that CMS electrodes with MgO tunnel barriers are important combinations for extending the application of Heusler alloy films to spintronic devices. Figure 4(b) shows the $\beta$ dependence of the Co spin magnetic moment at different sites deduced from the KKR-CPA calculation. While the magnetic moment of Co at the regular site $m_\mathrm{spin}$(Co$_\mathrm{Co}$) behaves similarly to the average Co moment, that of Co at the Mn site,$m_\mathrm{spin}$(Co$_\mathrm{Mn}$), is larger than the average Co moment and further increases with $\beta$. This is because, for large $\beta$, Co atoms at the Mn sites become dilute (Table I of ref. \cite{3} and behave like isolated antisite Co atoms, which show an enhanced magnetic moment. The decrease in the spin moment of Co at the regular site $m_\mathrm{spin}$(Co$_\mathrm{Co}$) is also attributed to the reduction in the majority-spin charge due to the delocalization of Co 3$d$ states with increasing $\beta$. This can be confirmed by the LDOSs of Co 3$d$ at the regular site \cite{3}.  It was also reported that the $m_\mathrm{spin}$ (Co) of Co-rich CMS/MgO was 1.1 $\mu_\mathrm{B}$ \cite{6} for a sample at room temperature which is consistent with our results. Here, the CMS film composition of Co: Mn: Si=2: 0.7: 1 is Co-rich similar to ref. \cite{6}. Consequently, the Co-rich CMS would contain more or less the same amount of Co$_\mathrm{Mn}$ as in Co-rich Co$_\mathrm{2}$Mn$_{\beta}$Ge$_\mathrm{0.38}$ (CMG) \cite{2}. Picozzi {\it et al.}\cite{7} reported that in-gap states could theoretically exist within the minority spin gap for CMS with Co$_\mathrm{Mn}$ . The spin polarization of Co-rich CMS estimated from the TMR ratio at 4.2 K for CMS/MgO MTJs assuming Julliere’s model, (P)CMS, was as low as 0.75 \cite{29}.

\begin{figure}[htbp]
\begin{center}
\includegraphics[width=09cm]{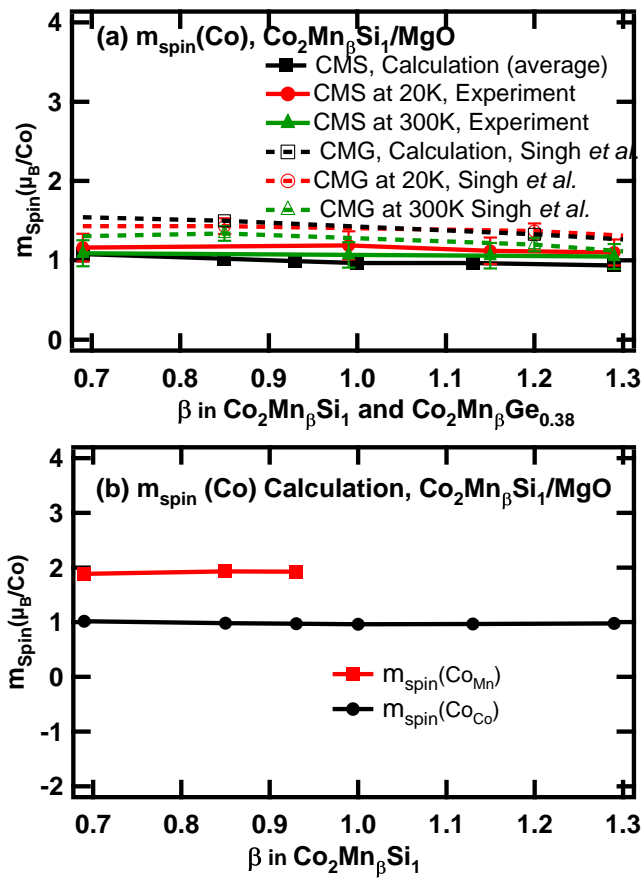}
\caption{(Color online) (a) Mn composition ($\beta$) dependence of the Co spin magnetic moment for Co$_\mathrm{2}$Mn$_{\beta}$Si$_\mathrm{1}$ deduced using the sum rules \cite{1,2,19,24,25,26,26A,26B,26C,26D} compared with that given by the present KKR-CPA calculation. The magnetic moments of Co in CMG by Singh  {\it et al.}\cite{19} are shown by dotted lines. (b)$\beta$ dependence of the Co spin magnetic moments at different sites deduced from present KKR-CPA calculation for Co$_\mathrm{2}$Mn$_{\beta}$Si$_\mathrm{1}$.}
\end{center}
\end{figure}

\ \ \ Comparison between Co-rich CMG and Co-rich CMS may be summarized as follows: (i) the deviation in composition from 2:1:1 for CMG is larger than that for CMS, (ii) the spin polarization at $E_\mathrm{F}$ (P)CMG of 0.74 \cite{30} is comparable to the (P) CMS of 0.75, \cite{29}  and (iii) the experimental $m_\mathrm{spin}$(Co) of the present XMCD results and the theories for CMS and CMG are in good agreement. Importantly, the creation of Co$_\mathrm{Mn}$ antisites in Mn-rich CMS electrodes would be suppressed because a Mn$_\mathrm{Co}$ antisite has much lower formation energy than a Co$_\mathrm{Mn}$ antisite. The suppression of the Co$_\mathrm{Mn}$ antisite formation would lead to a decreased density of in-gap states around $E_\mathrm{F}$  for Mn-rich CMS and thereby lead to decreased tunnel conductance for antiparallel (AP) spins. Therefore, we obtained higher TMR ratios for Mn-rich CMS electrodes up to 1995\% at 4.2 K and 354\% at RT \cite{5}. Yamamoto  {\it et al.} \cite{4} showed that the picture of suppressed Co$_\mathrm{Mn}$ antisites with increasing $\beta$ in CMS and CMG films is consistent with the SSFU composition models introduced by taking into account the theoretical formation energies of  various kinds of defects induced in Co$_\mathrm{2}$MnSi and Co$_\mathrm{2}$MnGe.  Furthermore, they showed that the $\beta$ dependence of the saturation magnetization per formula unit of CMS films,  $\mu_\mathrm{s}$, was well explained by the KKR-CPA calculations based on the antisite-based SSFU composition model \cite{3}. They also showed the $\beta$ dependence of  $\mu_\mathrm{s}$ of Ge-deficient CMG was qualitatively explained by the model \cite{4}. These findings also support the validity of the model for CMS and CMG films \cite{3,4,19}. Thus, our experimental findings are consistent with the scenario that the density of minority-spin in-gap states can be reduced by appropriately controlling defects in Co$_\mathrm{2}$MnSi and Co$_\mathrm{2}$MnGe electrodes. Off-stoichiometry is unavoidable, to various degrees, in Co$_\mathrm{2}$MnSi and Co$_\mathrm{2}$MnGe thin films which are mostly prepared by magnetron sputtering. To put it briefly, Co$_\mathrm{Mn}$ antisites, which are harmful to the half-metallicity of CMS, would be suppressed with an increasing Mn composition, resulting in a decreased density of in-gap states around $E_\mathrm{F}$ of CMS electrodes \cite{4,19}.

\section{Conclusion}

\ \ \  We have studied the magnetic and electronic states of the CMS/MgO interfaces by means of XMCD. The deduced $m_\mathrm{spin}$ (Mn) for all the CMS samples decreased with Mn composition ($\beta$), consistent with Li  {\it et al.}'s calculation and Picozzi  {\it et al.}'s calculation which predict that Mn atoms at both the regular sites and the Si sites lose the magnitudes of the spin moment with $\beta$. The deduced $m_\mathrm{spin}$(Co) values for $\beta<$1 in CMS samples are larger than the theoretical value of 1.05 $\mu_\mathrm{B}$ for bulk stoichiometric CMS because in the Co-rich region, the existence of Co$_\mathrm{Mn}$ antisites increases and Co$_\mathrm{Mn}$ has a larger magnetic moment, $m_\mathrm{spin}$(Co) $\sim$ 1.16$\mu_\mathrm{B}$, which on the other hand reduces the spin polarization at the Fermi level and decreases the TMR ratio. However, the deduced $m_\mathrm{spin}$ (Co) was slightly decreased with increasing $\beta$ for  $\beta>$1 and this is because in the Mn-rich CMS the amount of Co$_\mathrm{Mn}$ decreases compared to Co-rich CMG. The experimental results are consistent with the Li {\it et al.}'s calculation. Thus, our XMCD study suggests that CMS electrodes with MgO tunnel barriers are favorable combinations for extending the application of Heusler alloy films to spintronic devices compared to Ge-deficient CMG.

\begin{center}
{\bf Acknowledgement}
\end{center}

\ \ \  We would like to thank Kenta Amemiya and Masako Sakamaki for valuable technical support. This work was supported by a Grant-in-Aid for Scientific Research (S) (No. 22224005) from JSPS and the Quantum Beam Technology Development Program from JST, Japan. The experiment at Photon Factory was approved by the Program Advisory Committee (Proposal Nos. 2012G667, 2010G187 and 2010S2-001). The work at Hokkaido University was partly supported by a Grant-in-Aid for Scientific Research (A) (No. 23246055) from JSPS, Japan.

\end{document}